\documentclass[
    ,final            
  ]
  {aipproc}

\layoutstyle{8x11double}

\def\apj{ ApJ}
\def\aap{ A\&A}
\def\mnras{MNRAS}

\def\pasj{PASJ}

\def\apjl{ ApJL}

\begin{document}

\title{Formation of Millisecond Pulsars in Globular Clusters}


\author{Natalia Ivanova}{
  address={CITA, University of Toronto, 60 St George St, Toronto, ON M5S 3H8, Canada}
}

\author{Craig O. Heinke}{
  address={Department of Astronomy, University of Virginia, 530 McCormick Road
Charlottesville, VA 22904-4325, USA},
altaddress={Physics and Astronomy Department, Northwestern University, 2145 Sheridan Rd, Evanston, IL 60208 USA }
}

\author{Frederic A. Rasio}{
address={Physics and Astronomy Department, Northwestern University, 2145 Sheridan Rd, Evanston, IL 60208 USA }
}

\begin{abstract}
In this contribution we discuss how neutron stars are produced and retained
in globular clusters, outlining the most important
dynamical channels and evolutionary events that affect the
population of mass-transferring binaries with neutron stars
and result in the formation of recycled pulsars.
We confirm the importance of electron-capture supernovae
in globular clusters as the major supplier of retained neutron stars.
By  comparing the observed millisecond pulsar population and the results
obtained from simulations, we  discuss several constraints on the evolution
of mass-transferring systems.
In particular, we find that in our cluster model the following mass-gaining events create populations of MSPs that 
do not match the observations  (with respect to
binary periods and companion masses or the number of produced systems) and therefore likely do not
lead to NSs spun up to millisecond periods: (i) accretion during a common envelope event with a
NS formed through accretion-induced collapse, and (ii) mass transfer from a WD donor. 
By restricting ourselves to the evolutionary and dynamical paths that most likely lead to neutron star recycling,
we obtain good agreement between our models and the 
numbers and characteristics of observed millisecond pulsars in the clusters Terzan 5 and 47 Tuc.
\end{abstract}


\maketitle


\section{Introduction}

Millisecond pulsars (MSPs) are present in globular clusters (GCs) in great numbers: 
about 140 GC MSPs have been detected\footnote{See http://www.naic.edu/~pfreire/GCpsr.html for an updated list.},
with more than a dozen in several GCs -- in 47~Tuc \citep{Camilo00},
M28 \citep{Stairs06},  and Terzan~5 \citep{Ransom05}. 
It is estimated that of order 1000 potentially detectable MSPs are present in the Galactic GC system \citep{Heinke05a},
as at present the pulsar searches have reached the bottom of the pulsar luminosity function only in a few clusters (e.g., 47~Tuc and M15).

Per unit mass, the number of MSPs in GCs greatly exceeds their numbers in the Galaxy.
This was, however, an expected discovery: MSPs are thought to be descendants of LMXBs (for a review see, e.g., Bhattacharya and van den Heuvel 1991),
and the abundance of  X-ray binaries per unit mass was known to be $\sim$100 times greater in GCs than in the Galaxy as a whole \citep{Clark75}.
The high formation rate of X-ray binaries, in itself, is widely accepted to be a consequence of the high stellar density of GCs, 
which may lead to the creation of compact NS binaries in close stellar encounters.

Although the numbers of MSPs and X-ray binaries per unit mass in globulars are much higher than those in the rest of the Galaxy,
only a few X-ray binaries or a few dozen MSPs are present per fairly massive (more than a million stars) 
and dense cluster. This makes the problem computationally very challenging.
E.g., the target time for direct $N$-body methods to address the million-body problem is as far in the future as 2020 
(Hut 2006). In our studies, we use a modified encounter rate technique method, described in detail in 
\citet{Ivanova05_bf}, and with the updates described in \citet{Ivanova07}.

\section{Neutron Star Production}

In our studies we adopt that a NS can be formed as a result of either a core-collapse (CC) supernova,
which occurs after an iron core is formed, 
or an electron capture supernova (ECS). 
The latter, it has been argued, occurs when a degenerate ONeMg core reaches  $M_{\rm ecs}=1.38 M_\odot$;
its collapse then is triggered by  electron capture on $^{24}$Mg and $^{20}$Ne 
before neon and subsequent nuclear burnings start, and therefore before iron core formation 
\citep{Miyaji80_ecs, Nomoto84_ecs1, Nomoto87_ecs2,Timmes92_cotoonems,Timmes94_cotoonems}. 

There are several possible situations when a degenerate ONeMg core can be developed and
reach $M_{\rm ecs}$:

\begin{itemize}
\item {\it Evolutionary induced collapse} (hereafter EIC). 

If the initial core mass is less than required for neon ignition, $1.37 M_\odot$,
the core becomes strongly degenerate and grows to $M_{\rm ecs}$ through the continuing He shell burning. 
The critical mass range for this to occur in single stars is somewhere between 6 and 10 $M_\odot$.
In more massive stars, carbon, oxygen, neon and silicon burnings
progress under non-degenerate conditions, and, in less massive stars, ONeMg cores never form.
This critical mass range depends on the properties of the He and CO cores, which, in turn, 
are highly dependent on the mixing prescription (semiconvection, overshooting, rotational mixing, etc.)
and varies between different evolutionary codes \citep[see discussion in][]{Podsi04_aic,Siess06_agb}.
In the code that we use for our cluster simulations, a non-degenerate ONeMg core
is formed when the initial He core mass is about $2.25\ M_\odot$ \citep{Pols98_models,Hurley00} and 
the ranges of initial masses for single stars that lead to the formation
of such a core are  6.85 to 7.57  $M_\odot$ and  6.17 to 6.76  $M_\odot$
for metal-rich GCs ($Z=0.005$) and  metal-poor GCs ($Z=0.0005$), respectively. 
In binaries, the mass transfer history of the star may affect the range of progenitor masses
for which an ECS can occur \cite{Podsi04_aic}. We find that in our simulations 
the mass range can extend to as low  as 3  $M_\odot$ and to as high as 22  $M_\odot$.

\begin{table}
\begin{tabular}{@{}l  c c c c }
\hline
 & CC & EIC & AIC & MIC   \\
\hline
single \\
 Z=0.0005& 3354 & 594& - & -  \\
 Z=0.001  & 3255 & 581 & - & -  \\
 Z=0.005 & 2833 & 570 & - & -  \\
 Z=0.02 & 2666 & 400 & - & -  \\
binary \\
 Z=0.0005  & 3079 & 545 & 59 & 14 \\
 Z=0.001  & 3056 & 553 & 60 & 15 \\
 Z=0.005  & 2750 & 576 & 58 & 16  \\
 Z=0.02  & 2463 & 406 & 33 & 20  \\
\hline
\end{tabular}
\caption{Production of NSs.}
\tablenote{Production of NSs in the case of no dynamics.
Notations for channels -- CC - core-collapse supernova; EIC - evolution-induced collapse;
AIC - accretion induced collapse; MIC - merger-induced collapse.
Numbers are scaled per 200,000 $M_\odot$ stellar population mass at the age of 11 Gyr.}
\label{tab-ns-nod}
\end{table}

\item{\it Accretion induced collapse} (hereafter AIC). 

AIC can occur by accretion on to a degenerate ONeMg WD in a binary, where
such a WD steadily accumulates mass until it reaches the critical mass $M_{\rm ecs}$.
The set of conditions for the MT rates which allow mass accumulation can be found in
\citet{Ivanova04_ttmt}.

\item{ \it Merger induced collapse} (hereafter MIC).

When two WDs coalesce and the product has a total mass
exceeding $M_{\rm ecs}$, either one of the WDs is a massive ONeMG WD
or the two are both CO WDs. In the latter case, 
 an off-centre carbon ignition converts the coalesced star into an ONeMg core,
and then the WD proceeds with an ECS as usual \citep{Saio85_cowd,Saio04_cowd}.

\end{itemize}

The gravitational mass of the newly formed NS is $\sim 0.9$ of its 
baryonic mass and will be $1.26\ M_\odot$ in the case of AIC or EIC. 
However, in the case of MIC, it is possible to form a more massive NS.
The reason is that the condition for ECS to occur depends on the central density. 
As a consequence, the collapse of a rapidly rotating WD, which was formed during the coalescence of two WDs 
with total mass exceeding the Chandrasekhar mass,
can lead to the formation of a more massive, and also rapidly spinning, NS.

In Table~\ref{tab-ns-nod} we show the results for NS production via different channels
in  stellar populations of several metallicities -- from solar (as in the Galactic field) to that of a typical 
metal-poor  cluster: Z=0.02, 0.005, 0.001, 0.0005. 
Note that the production of NSs via core-collapse SNe (CC NSs), per unit of total stellar mass 
at 11 Gyr, decreases as metallicity increases. 
The difference between the solar metallicity case and the metal-poor case is about 20 percent, while the difference 
between metal-rich and metal-poor clusters (Z=0.005 and Z=0.0005) is about 10-15 percent.

EIC in single stars for metal-rich populations occurs in stars of higher masses.
The range of initial masses is, however, the same for both populations and is about $0.6 M_\odot$.
As a result, in agreement with the adopted Kroupa power law for the IMF \citep{Kroupa02}.
the number of NSs produced via EIC (EIC NSs) from the population of single stars is smaller
in the case of a Galactic field population than in the case of a GC population by 30 per cent,
while the difference between metal-rich and metal-poor clusters is only 5 per cent. 

AIC and MIC NSs can be produced only in populations with binaries.
The range of initial masses of progenitors is not as strongly defined as for EIC in single stars.
As a result, their production rates do not have  a strong dependence  on the metallicity.

\section{Neutron Star Retention}

To estimate how many NSs can be retained in GCs,
we first considered stellar populations without dynamics.
In our simulations, we adopt the most recently derived pulsar kick velocity distribution 
from \cite{Hobbs05_kicks}. It is a Maxwellian distribution with one-dimensional RMS velocity
$\sigma = 265$ km s$^{-1}$.
For comparison, we also considered a case with the natal kick distribution as in  earlier studies \citep{Arzoumanian02ApJ_kicks}
(with Z=0.005, typical for a metal-rich globular cluster). There, 
the kick velocity distribution is 2 Maxwellians with a lower peak for kick velocities at 90 km/s.

\begin{figure}
  \includegraphics[height=.3\textheight]{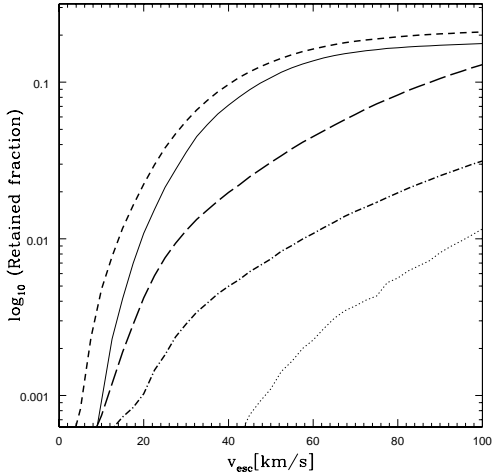}
 \caption{The retention fractions as a function of escape velocity (for stellar evolution 
unaffected by dynamics) for a \cite{Hobbs05_kicks} kick distribution. 
Dotted and dash-dotted lines show the retention fractions for single and binary 
populations,
core-collapse NSs only. Solid and short-dashed lines show the total retention 
 fractions for single and binary populations, all NSs. 
For comparison, we show  the total retention fraction of a binary 
population with the \cite{Arzoumanian02ApJ_kicks} kick distribution (long-dashed line).}
\label{fig:ret}
\end{figure}
 
For ECS NSs from all the channels (EIC, AIC and MIC), we adopt that the accompanying natal kick is
10 times smaller: we adopt the results of numerical simulations which find 
that the SASI instability, required by current understanding for the large explosion asymmetry in the case of core-collapse supernovae,
fails to develop \cite{Buras06,Kitaura06}). As a result, natal kicks for ECS NSs do not exceed 100 km/s.

We find that even considering a stellar population with 100\%  primordial binaries,
the retention fraction of CC NSs is very small (Fig.~\ref{fig:ret}); 
the resulting number of retained CC NSs is just a few per  typical dense globular cluster of $2\times10^5 \ M_\odot$.
In contrast, NSs formed via different ECSs channels provide about 200 retained NSs per typical GC (similar numbers were found also in \cite{Kuranov06}).
Therefore, in contrast to the population of NSs
in the Galaxy, the population of NSs in GCs is mainly low-mass NSs made by ECS. 
With the old distribution for natal kick velocities, reasonable numbers of NSs can be produced and
retained in a typical GC, while a massive cluster like 47~Tuc could retain as many as 600 NSs 
(see also Fig.~\ref{fig:ret}). This result agrees with the retention fractions obtained in \citet{Pfahl02}.

In the case when dynamical encounters affect the stellar population, and, as an example,
 could destroy primordial binaries, the resulting numbers of produced and retained NSs
are somewhere between only-primordial-single star case and only-primordial-binary case.
The exceptions are AIC and MIC NSs, the formation of which is enhanced by dynamical encounters.
Overall, a typical dense cluster will retain 8 CC NSs, $\sim$170 EIC NSs, $\sim$50 AIC NSs
and $\sim$20 MIC NSs; Terzan 5 will retain in total about 500 NSs (assuming that the cluster mass
is 370 000 $M_\odot$) and about 1100 NSs will be in a cluster like 47 Tuc (if its total mass is $10^6$ $M_\odot$).

\section{Neutron Star Recycling}

\begin{figure}
\includegraphics[height=.35\textheight]{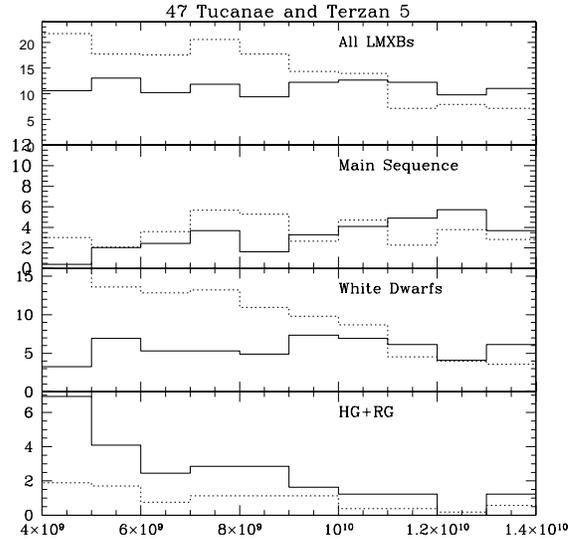}
\caption{
Number of appearing LMXBs per Gyr with different donor types (with all donors, with main sequence donors, with white dwarf donors and with 
donors that are red giants or subgiants or are in the Hertzsprung Gap) for 47 Tuc (solid line) and Terzan 5 (dotted line).}
\label{fig-lmxb-zoo}
\end{figure}

A common understanding of MSP formation is that the NS is recycled through disk 
accretion. The formation rate of LMXBs can therefore be linked with the number of observable MSPs. 
The connection between the formation rate of LMXBs and their existence at a give time in a GC is provided by the life-time $\tau_{\rm LMXB}$ of an LMXB with a particular donor.
An average $\tau_{\rm LMXB}$ for NS-MS LMXBs is about 1 Gyr. Depending on the metallicity and the initial donor mass,
a system can be persistent 5-40 percent of the MT time for metal-rich donors $>0.6 M_\odot$ and
transient at the rest;
for donors of lower metallicities or smaller masses, a NS-MS LMXB will be transient all the time \citep{Ivanova06_lmxb},
and therefore most likely be seen as a quiescent LMXB (qLMXB) rather than a bright LMXB.
In the case of NS-WD LMXBs (ultra-compact X-ray binaries, UCXBs), total $\tau_{\rm LMXB}$ is few Gyr,
however the time when a system is persistent and has an X-ray luminosity above $10^{36}$ erg/s is $10^7\div10^8$ yr only.
An LMXB with a red giant companion or a companion that is in the Hertzsprung gap is very short-lived,
$10^5-10^7$ yr, and in only very rare cases can they live as long as $10^8$ yr.

In our simulations, we find that a typical GC can contain up to
2 LMXBs with a MS companion (most likely observed, at any particular time, as qLMXBs) and up to one LMXB with
a WD companion  (ultra-compact X-ray binaries, UCXBs). The scatter in the average number of observed LMXBs per cluster in  
independent simulations is rather large (more than expected from Poisson statistics) - e.g., for UCXBs, it can vary between 0.1 and 1.1. 
In the case of Terzan 5 and 47 Tucanae, the average number of LMXBs formed per Gyr, at the 
age of 11 Gyr, is $\sim 5$ for NS-MS LMXBs and $\sim 8$ for UCXBs (see Fig.~\ref{fig-lmxb-zoo}). 
These numbers are in general agreement with the observations: 
(i) assuming that $\tau_{\rm UCXB}\sim10^8$ yr, those numbers are consistent with  the presence of one bright UCXB in Terzan 5 
and no bright UCXB detected in 47 Tuc; (ii) several qLMXBs are identified in both Terzan 5 and 47 Tuc.

Overall the numbers of NSs that gain mass via mass transfer (MT) through 11 Gyr of cluster evolution
are high: for our 47 Tuc model, about 50 NS-MS binaries and about 100 UCXBs.
As we observe fewer MSPs in these GCs, while the rate of LMXB formation in simulations 
is consistent with the observations, we conclude that not all NSs that gain mass via MT become currently active MSPs.

Indeed, it has been argued that a NS is spun up only if the accretion rate is not too low, $\dot M \ge 3\times 10^{-3}\dot M_{\rm Edd}$,
where $ \dot M_{\rm Edd}$ is the Eddington limit (for a review see, e.g., Lamb \& Yu 2005).
In a UCXB, soon after the start of mass transfer,
the accretion rate drops very quickly.  After 1 Gyr,
it is less than $10^{-4} \dot M_{\rm Edd}$. Such a MT leads to a spin-down of the previously spun-up NS, and no MSP is formed. 
Support for this statement is given by the fact that no 
UCXBs (those that have WD companions) are visible as radio MSPs (Lamb \& Yu 2005).    

The requirement of steady spin-up through disk accretion implies that not all physical collisions
will lead necessarily to NS spin-up. In the case of a physical collision with a giant,
the NS will retain a fraction of the giant envelope, with a mass of a few hundredths of $M_\odot$ (Lombardi et al. 2006).
Immediately after the collision, this material has angular momentum and most likely will form a disk. 
We adopted therefore that in the case of a physical collision with a giant, an MSP can be formed, but, 
in the case of any other physical collision, the NS will not be recycled.

\section{Millisecond pulsars}

\begin{figure}
\includegraphics[height=2.5in,width=2.5in]{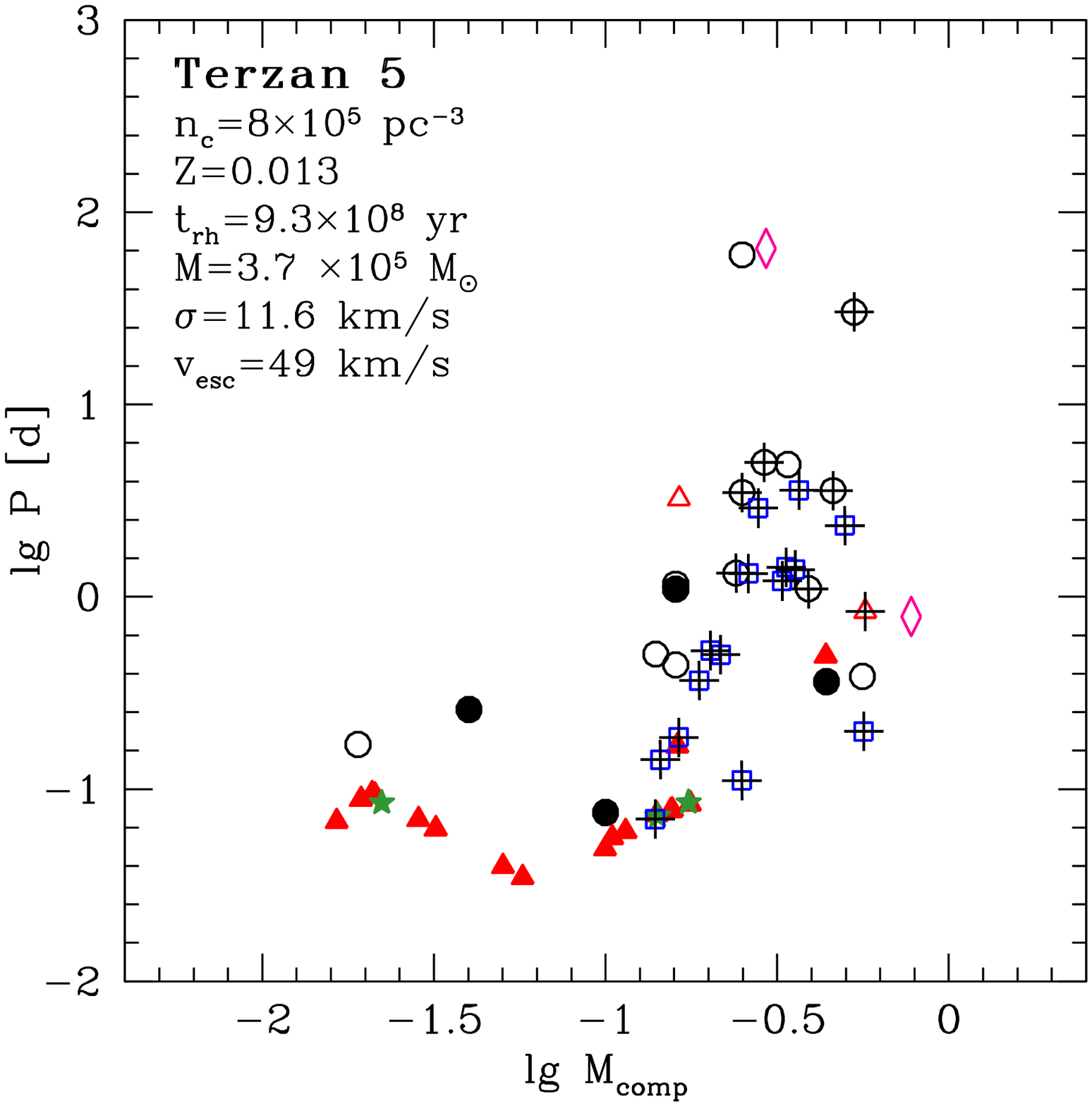}
\includegraphics[height=2.5in,width=2.5in]{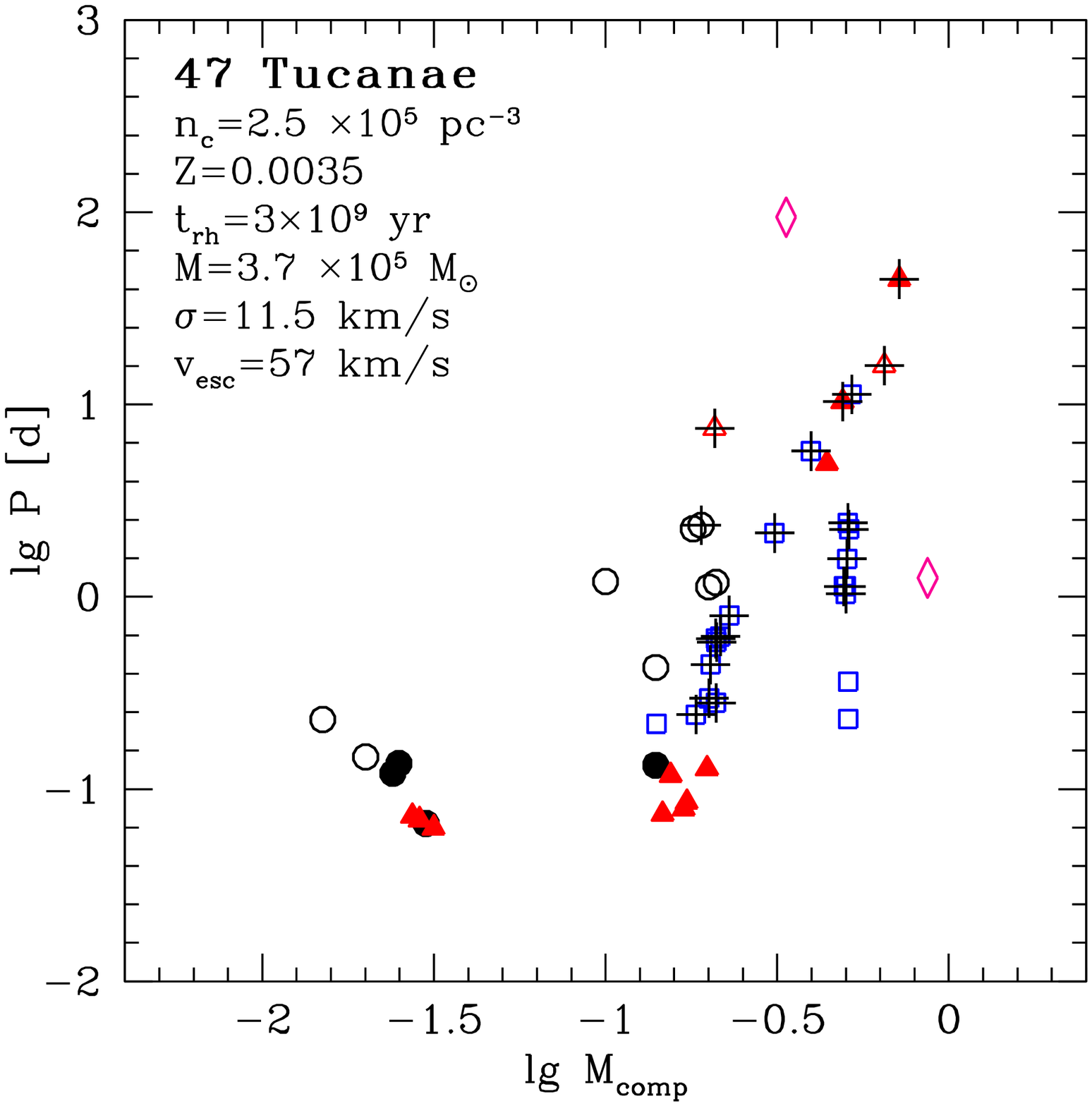}
  \caption{bMSPs in simulated models of 47 Tuc and Terzan 5 compared to observed bMSPs.  The simulated populations correspond to several independent runs and represent a larger 
population than in the observed clusters. Observed bMSPs are shown with circles;
triangles - bMSP formed via binary exchanges; stars - via tidal captures;
squares - via physical collisions; diamonds - primordial binaries. 
 Cross signs mark eccentric bMSPs ($e\ge0.05$) and solid symbols mark systems with a non-degenerate companion (in the case of simulations) or observed eclipsing systems.  
}\label{fig:msp}
\end{figure}

Suppose that all mass-gaining events in the life of a NS -- mass transfer,
physical collision with a red giant, common envelope hyper-accretion or merger --
can lead to NS recycling. 
In this case we found that as many as 250 and 320 potential MSPs
are made in our simulations of clusters like Terzan 5 or 47 Tuc, respectively 
(the corresponding numbers of retained NSs are $\sim 500$ and $\sim 1100$).
Although these numbers correlate well with the formation rate of LMXBs, 
they greatly exceed the numbers of observed and inferred MSPs in both clusters, 
which are 33 (in Terzan 5; perhaps 60 total) and 22 (in 47 Tucanae; perhaps 30 total).

By considering the whole population of NSs that gained mass,
we found that bMSPs formed from primordial binaries, where a common envelope event led to AIC,
create a population of potential bMSPs with relatively heavy companions, 
in circular orbits with periods from one day to several hundreds of days. 
This population is not seen in either Terzan 5 or 47 Tuc. 
We considered primordial binaries that evolved through  
mass transfer from a giant donor after a NS was formed via AIC.  
Even though bMSPs that have similar periods, companion masses and eccentricities
are present in Terzan 5, there are no such systems in 47 Tucanae.
Also, bMSPs made from primordial binaries after AIC must inevitably be formed in low-dense clusters, 
but no bMSPs are observed there. These facts tell us that either 
AIC does not work, or the kicks in the case of AIC are stronger then we adopted, 
or a NS formed via AIC has such a strong magnetic field, that surface accretion does not occur.

Let us summarize the mass-gain scenarios described above (in this and the previous Section)   
which might not lead to the formation of MSPs:

\begin{itemize}
\item{Primordial bMSPs where a NS gained mass via CE}
\item{bMSPs where a NS gained mass via merger}
\item{Mass transferring NS-MS systems with a donor mass above $\sim0.05-0.1M_\odot$, 
as they most likely are seen only as LMXBs/qLMXBs}
\item{Mass transfer in NS-WD systems does not produce millisecond pulsars}
\end{itemize}

Considering all the exclusions described above, we form in our simulations at least   
$15\pm7$ MSPs for Terzan 5 and $25\pm4$ MSPs for 47 Tuc (for the formed population of bMSPs, see Fig.~3).  
The values for Terzan 5 are somewhat uncertain due to uncertainty in the properties of this heavily reddened cluster.
The binary population of MSPs is $1/2-1/3$ of all MSPs.
The total number of NSs that gain mass in the simulations are 250 and 320
in Terzan 5 or 47 Tuc, accordingly.

We predict that the fraction of single pulsars is higher in Terzan 5 than in 47 Tuc, as observed.
The origin of isolated MSPs depends on the cluster dynamical properties.
For example, in our standard model, about half of isolated MSPs 
were formed in a result of an evolutionary merger at the end of the MT from a MS companion,
and another half lost their companion as a result of a binary encounter.
Most such binary encounters occurred in systems where a NS was spun-up during  MT from a giant,
and had a WD companion just before it become single.
In the case of 47 Tuc, very few isolated pulsars were formed in a result of 
an evolutionary merger at the end of the MT. Most lost their companion as a result
of a binary encounter, where in half of the cases the companion before the binary destruction
was a low-mass WD or MS star at the end of their MT sequence.

MSPs are likely to be located where they were formed.
Most of them are in the core and less than a third in the halo.
Halo MSPs cam be primordial or recoiled, where the 
fraction of recoiled MSPs increases as  $v_{\rm rec}$ decreases.
Roughly half of observed pulsars are found outside their cluster core radius \citep{CamiloRas05}, 
but the radial distribution of most pulsars is exactly as expected for a population 
that is produced in or around the core \citep{Grindlay02, Heinke05a}, 
with the exception of a few, likely ejected, pulsars in M15 and NGC 6752 \citep[e.g.][]{Colpi02}.

\section{Discussion}

We considered in detail the problem of the formation of MSPs in GCs,
which includes NS formation, retention and recycling.
Our simulations showed that most of the retained neutron stars in GCs
must come from an electron capture supernova formation channel (c.f. the field,
where most NSs are from core-collapse supernovae). 
A typical GC could contain at present (where the adopted cluster age is 11 Gyr) 
as many as $\sim 200$ NSs where about half of them 
are located outside the core; a massive GC like 47 Tuc could have more than a thousand NSs.

Our simulations produce LMXBs in numbers comparable to observations if indeed
as many as a few hundred NSs are retained per GC. 
If, contrary to our assumptions, AIC does not lead to the formation of NSs,
then the number of formed NSs is reduced only by $\sim 20\%$, but the number
of appearing LMXBs is decreased by 2-3 times (per Gyr, at the cluster age of 11 Gyr).
Given the large scatter in the simulations, it may still be consistent with the observations.

We find that, in case of a very dense cluster, up to half of NSs could gain mass after their formation through
mass transfer, hyper-accretion during a common envelope, or physical collision. 
It is likely that most of these mass-gaining events do not lead to  NS spin-up, 
and that only a few percent of all NSs appear eventually as MSPs, 
implying that there is a large underlying population of unseen NSs in GCs. 
In particular, we find that if we assume that a NS would accrete and spin up during a common envelope, 
we overproduce bMSPs in the cluster halos from primordial binaries of intermediate masses.
Such bMSPs would be present in low-density clusters and have not yet been seen.

Excluding the systems discussed above, as well as those which are still actively accreting their donor's material 
and are seen instead as LMXBs, we obtained the number of ``detectable'' bMSPs.
The predicted numbers -- about 2 dozen -- are in good agreement with the observations.
The fraction of isolated pulsars is comparable with that observed and is bigger in Terzan 5 than in 47 Tuc. 
Comparing the population census of our models with the observations  of all detected pulsars to date in GCs,
we find good agreement for all types of pulsars.

In conclusion we outline several important questions that must be addressed
for further progress in studies of NSs in globular clusters: 

\begin{itemize}

\item What is the final fate of a mass-transferring NS-WD binary --
the post-MT period, the NS luminosity in its quiescent state, and how does it lose its companion?

\item What is the result of mass accretion on a NS during  
a merger or a physical collision?

\item For a common envelope event in primordial binaries of intermediate masses 
that produce a NS via ECS -- does the NS accrete the material 
and spin up?  What is the common envelope efficiency?

\end{itemize}

\bibliographystyle{aipproc}   

\end{document}